\documentclass[twocolumn]{aastex631}
\usepackage{amsmath,natbib}

\newcommand{\Msun}{\,{\rm M_\odot}}

\newcommand{\Mblack}{M_\bullet}

\begin{document}

\title{General Relativistic Shock Wave Solutions with Black Hole Formation: \\The Singular Isothermal Sphere Case}

\author[0000-0002-4945-5079]{Chien-Ting J. Chen}
\affiliation{Science and Technology Institute, Universities Space Research Association, Huntsville, AL 35805, USA}
\affiliation{Astrophysics Office, NASA Marshall Space Flight Center, ST12, Huntsville, AL 35812, USA}

\author{Michael J. Cai}
\affiliation{Google, 1600 Amphitheatre Pkwy, Mountain View, CA 94043, USA}

\author[0000-0001-9879-7780]{Fabio Pacucci}
\affiliation{Center for Astrophysics, Harvard \& Smithsonian, 60 Garden St, Cambridge, MA 02138, USA}
\affiliation{Black Hole Initiative, Harvard University, 20 Garden St, Cambridge, MA 02138, USA}

\begin{abstract}
The rapid emergence at $z\gtrsim 6$ of ubiquitous populations of supermassive black holes (SMBHs) revealed by JWST and of quasars with estimated masses $\Mblack > 10^{10} \Msun$ demands efficient pathways for early growth. The smooth collapse of a singular isothermal sphere (SIS) has been solved analytically in full general relativity, but the shock waves that inevitably accompany such collapse have not. Here, we derive general-relativistic self-similar shock-wave solutions for the collapse of an SIS to a black hole, extending the framework of \citet{cai05} to discontinuous flows. We obtain the general relativistic jump conditions for an isothermal fluid and show that they connect interior collapse solutions to exterior envelopes that may be static, expanding, or collapsing, yielding a rich family of shocks propagating at up to $\sim$40\% the speed of light; the available exterior types narrow with increasing sound speed. A coordinate-matching technique that uses the zero-velocity surface uniquely bridges the Schwarzschild and comoving self-similar descriptions, completing the characterization of the growing black hole. The central accretion rate is set by the interior collapse alone and is suppressed by a factor of $\sim$5--7 relative to the smooth expansion-wave solution, while the energy released at the shock reaches $\sim$10\% of the enclosed rest mass---nearly twice the 5.7\% radiative efficiency of Schwarzschild accretion. These results provide an analytical energy budget for direct-collapse black hole formation, with implications for SMBH seed assembly, the dense cocoons around nascent high-redshift black holes, the recently discovered JWST's Little Red Dots, and relativistic transients such as gamma-ray bursts.
\end{abstract}


\section{Introduction} \label{sec:intro}

JWST has revealed an unexpectedly abundant population of active black holes at the earliest cosmic epochs. Broad-line AGN surveys have identified accreting SMBHs with masses $\sim 10^{6}$--$10^{8}\,M_\odot$ out to $z\sim 4$--$7$ \citep[e.g.,][]{Lai2023, Harikane2023, Kocevski_2023, Pacucci_2023_JWST, Kocevski_2024}, while the detection of a vigorously accreting $\sim 10^{6}\,M_\odot$ black hole in GN-z11 at $z=10.6$ \citep{Maiolino2024} and an X-ray luminous AGN at $z\approx 10.1$ \citep{Bogdan2024} demonstrate that massive black holes were already in place within 400--500~Myr of the Big Bang. Furthermore, the recent discovery of abundant ``Little Red Dots'' (LRDs) at $z \sim 5$ implies a vast population of obscured, massive black holes whose dense accretion envelopes might be the result of super-Eddington growth or late-stage quasi-star collapse \citep{Matthee2024, Pacucci_Narayan_2024, Begelman2026, Pacucci_2026_DCBH}. These discoveries have intensified a long-standing puzzle in black hole astrophysics: how supermassive black holes (SMBHs) assembled so rapidly and abundantly. Standard models of Eddington-limited accretion starting from stellar-mass seeds fall short by orders of magnitude in the available time \citep{Inayoshi2020, Volonteri2021}.

Several alternative pathways have been proposed. The direct collapse of primordial gas clouds in metal-free atomic cooling halos can produce ``heavy seeds'' of $\sim 10^{4}$--$10^{5}\,M_\odot$ \citep{Loeb_Rasio_1994, BL01, Bromm2003, Begelman2006, LatifFerrara2016}, a process that may be triggered by rapid gas accumulation and shock-heating during high-velocity collisions of protogalaxies \citep{Inayoshi2015}. Alternatively, runaway mergers in dense Population~III star clusters or rapid hierarchical growth from stellar-mass seeds have been explored \citep{PZ_2002, Katz_2015, Boekholt_2018, Woods2019, Regan2020}. While substantial progress has been made with numerical simulations, the general relativistic framework governing the dynamics of black hole formation through gravitational collapse---and the accompanying energy release---remains analytically underexplored. Yet this energy release is of direct astrophysical interest: a collapse that deposits a substantial fraction of the rest-mass energy into the surrounding gas could power the thermal emission \citep{DiMatteo_2008} and inflate the dense cocoons now inferred around the youngest accreting black holes \citep{Pacucci_2026_DCBH}, and regulate the rate at which the seed is fed.

The singular isothermal sphere (SIS) provides a powerful yet tractable model for self-gravitating collapse. In Newtonian gravity, \citet{shu77} found the celebrated expansion wave solution (EWS) for inside-out collapse, which has become the standard framework for low-mass star formation and has been confirmed in numerous observations of protostellar infall profiles. The subsequent discovery by \citet{Tsai1995} of shock wave solutions propagating at $\sim$1.26 times the sound speed. \cite{Lou2004} further extended the solution space to include envelope expansion with core collapse (EECC), and \citet{Shu2002} applied similar techniques to model champagne flows in H~II regions. Numerical simulations of the first star formation in a cosmological context \citep{Abel2002, BrommYoshida2011} have also demonstrated that primordial gas contracts roughly isothermally via atomic and molecular cooling, providing a direct physical motivation for extending isothermal collapse models to the general-relativistic regime. 

Building on CS05 \citep{cai05}, who formulated the GR self-similar collapse of the SIS in Schwarzschild and comoving coordinates and found the GR analog of Shu's EWS, we take the next analytical step---just as the Newtonian program advanced from the smooth EWS to shock wave solutions---by deriving the self-similar shock waves generated during black hole formation. \citet{Lian2014} treated dynamic polytropic collapse in a pseudo-Newtonian approximation; here we work in full general relativity, providing an analytical testbed for benchmarking general relativistic hydrodynamics (GRHD) codes in the strong-field regime \citep[e.g.,][]{Font2003,Rezzolla2013}.


The paper is organized as follows. Section~\ref{sec:formalism} briefly summarizes the self-similar framework of CS05. Section~\ref{sec:jump} derives the new shock jump conditions. Section~\ref{sec:solutions} presents the shock wave solutions. Section~\ref{sec:bh} describes the black hole properties using the comoving coordinate and the coordinate-matching technique. Section~\ref{sec:physics} discusses the mass accretion rate and shock energetics, and Section~\ref{sec:discussion} places the results in astrophysical context.

\section{Self-Similar Framework} \label{sec:formalism}

We adopt the self-similar Schwarzschild (SSS) formulation of CS05. The spherically symmetric metric is
\begin{equation}\label{eq:Smetric}
ds^2 = -\alpha^2 dt^2+a^2dr^2 + r^2\left(d\theta^2+\sin^2\theta\, d\phi^2\right),
\end{equation}
where $\alpha = \alpha(\zeta)$ and $a = a(\zeta)$ depend only on the self-similar variable $\zeta \equiv r/t$. The fluid obeys an isothermal equation of state, $P = \gamma\rho$, where $\gamma \equiv c_s^2$ is the squared sound speed.

In the orthonormal tetrad basis, the three-velocity $v$ measured by a local static observer defines the dimensionless energy density and velocity functions \citep[see CS05 for detailed derivations]{cai05}:
\begin{equation}\label{eq:e_nd}
\varepsilon=4\pi r^{2}(1+\gamma)\rho a^{2}, \quad \beta=-\frac{2v}{1-v^{2}}.
\end{equation}
The ``proper self-similar variable'' is $x^{2} \equiv a^{2}\zeta^{2}/\alpha^{2}$, and the hydrodynamic equations reduce to a set of ODEs:
\begin{subequations}\label{eq:ode}
\begin{alignat}{2}
x^{\prime} & = x-x\varepsilon \sqrt{\beta^2+1}-\beta\varepsilon \\
\beta^{\prime} & = \nonumber\\
-&\left\lbrace\varepsilon\beta\Gamma\left[\beta\left(x+\frac{1}{x}\right)+2\sqrt{\beta^2+1}\right]+2x(\varepsilon-1+\Gamma)D\right\rbrace \nonumber\\
& \times \left\lbrace\frac{x^2-1}{\sqrt{\beta^2+1}} + \left(\frac{2\beta x}{\sqrt{\beta^2+1}}+ x^2+1\right)\Gamma\right\rbrace^{-1} \\
\varepsilon^{\prime} & =  -\varepsilon^2\frac{\beta}{x}\left(2-\frac{\Gamma}{D}\right)
-\frac{\beta^{\prime}\varepsilon}{D}\left(\frac{1}{x}+\frac{\beta}{\sqrt{\beta^2+1}}\right) \\
\ln a^{2\prime} & = -\frac{\beta \varepsilon}{x},
\end{alignat}
\end{subequations}
where primes denote $d/d\ln\zeta$, $\Gamma\equiv(1-\gamma)/(1+\gamma)$, and $D\equiv\beta/x+\sqrt{\beta^{2}+1}+\Gamma$. Equation~\ref{eq:ode}d is immediately integrable: $a^{2}=1+\varepsilon D$. We refer readers to CS05 for the full derivation from the Einstein field equations.

\subsection{Critical Curves and Solution Strategy}\label{sec:critical}

The denominator of Eq.~\ref{eq:ode}b vanishes along a critical curve in the $x$--$\beta$ plane. Smooth passage requires simultaneous vanishing of the numerator, yielding the critical conditions (CS05):
\begin{equation}\label{eq:crit_xep}
x=\frac{-\beta\Gamma+\sqrt{1-\Gamma^{2}}}{1+\Gamma\sqrt{\beta^{2}+1}}, \quad
\varepsilon=1-\Gamma-\beta\Gamma\sqrt{\frac{1-\Gamma}{1+\Gamma}}.
\end{equation}
Applying l'H\^{o}pital's rule gives $\beta^{\prime}$ on the critical curve:
\begin{equation}\label{eq:crit_beta}
\beta^{\prime} = \frac{-m\pm\sqrt{m^{2}-4nl}}{2l},
\end{equation}
where $l$, $m$, and $n$ are functions of $\beta$, $x$, $\varepsilon$, $\Gamma$, and $D$ detailed in CS05 for displaying the equation in a compact form.

Given a choice of $\beta_{\rm crit}$, the values of $x$, $\varepsilon$, and $\beta^{\prime}$ on the critical curve are fully determined by Equations~\ref{eq:crit_xep}--\ref{eq:crit_beta}. This provides initial conditions for numerical integration of Eq.~\ref{eq:ode} via a fourth-order Runge-Kutta method. The exterior asymptotic boundary condition---a static SIS with $\beta_0=0$ and $\varepsilon_0=1-\Gamma$---or a hydrodynamic envelope with constant asymptotic $\beta_0$ and $\varepsilon_0$ can be matched via power series expansions in $1/x$ (see CS05, their Eq.~2.18).

\section{General Relativistic Shock Jump Conditions} \label{sec:jump}
The expansion wave solution (EWS) of CS05 passes through the critical curve at $\beta = 0$ and represents a smooth, continuous collapse. It is the unique continuous solution connecting the equilibrium exterior to a collapsing interior. All other initial conditions on the critical curve with $\beta_{\rm crit}>0$ yield solutions that we term ``collapse solutions with critical points'' (CSWCPs): these pass the critical curve smoothly in the first quadrant ($\beta>0$, collapsing), cross $\beta=0$, and then approach the critical curve again in the fourth quadrant ($\beta<0$, expanding). These CSWCPs cannot rejoin the exterior solution continuously---a shock discontinuity is required.

The physical reason is thermodynamic: while entropy increases as gas collapses inward through the critical surface, it would decrease if the gas were to expand back through it. The second law of thermodynamics therefore demands a shock or weak discontinuity at the critical surface for the expanding branch.

\subsection{Derivation of Jump Conditions}

For a fluid moving radially through a discontinuous surface, the general relativistic jump conditions in the shock rest frame (SRF) are \citep{LandauLevDavidovich1959}:
\begin{equation}\label{eq:jump_general}
\left[n\,v^{r}\right] = 0, \quad
\left[T^{tr}\right] = 0, \quad
\left[T^{rr}\right] = 0,
\end{equation}
representing conservation of baryon number flux, energy flux, and momentum flux across the shock, respectively. For an isothermal equation of state, the energy generated at the shock front is assumed to be radiated away immediately, so the energy flux condition is dropped. The remaining two conditions, written in components, are:
\begin{subequations}\label{eq:jump}
\begin{align}
&\rho_{\rm post} v_{\rm post} \gamma_{\rm post} = \rho_{\rm pre} v_{\rm pre} \gamma_{\rm pre} \label{eq:jump_num}\\
&(\rho_{\rm post}+P_{\rm post}){v_{\rm post}}^2{\gamma_{\rm post}}^2+P_{\rm post} \nonumber\\
&\qquad = (\rho_{\rm pre}+P_{\rm pre}){v_{\rm pre}}^2{\gamma_{\rm pre}}^2+P_{\rm pre}
\label{eq:jump_mom}
\end{align}
\end{subequations}
where the subscripts ``post'' and ``pre'' denote postshock and preshock quantities, $\gamma_{\rm post, pre}=(1-v_{\rm post, pre}^{2})^{-1/2}$ are the Lorentz factors in the SRF, and we have replaced the number density by $n=\rho/\mu$ (rest energy density divided by rest mass per particle).

The Landau-Lifshitz conditions are formulated in flat spacetime. In our tetrad formalism, the basis vectors are by definition orthonormal, thereby guaranteeing local flatness at the shock front. A Lorentz transformation connects the SRF to the SSS tetrad frame with the boost matrix:
\begin{equation}
\begin{pmatrix}
  \gamma_s & -\gamma_s v_s & 0 & 0 \\
  -\gamma_s v_s & \gamma_s & 0 & 0 \\
  0 & 0 & 1 & 0\\
  0 & 0 & 0 & 1
  \end{pmatrix},
\end{equation}
where $v_s$ is the shock front speed in SSS and $\gamma_s=(1-v_s^2)^{-1/2}$. Since $x \equiv \sqrt{g_{rr}}\,r/(\sqrt{-g_{tt}}\,t)$, the value of $x$ at the shock location gives the proper expansion speed: $v_s = x_s$.

Transforming Eq.~\ref{eq:jump} to the SSS frame yields:

\noindent\emph{\rm 1: Number flux conservation:}
\begin{equation}\label{eq:jump_sss_n}
\rho_{\rm post}\gamma_{\rm post} (v_{\rm post} - x_s) = \rho_{\rm pre}\gamma_{\rm pre} (v_{\rm pre} - x_s)
\end{equation}

\noindent\emph{\rm 2: Momentum flux conservation:}
\begin{equation}\label{eq:jump_sss_m}
\begin{aligned}
&\rho_{\rm post}\left[x_s^{2}{\gamma_{\rm post}}^{2}(1+\gamma)-x_s^{2}\gamma+{v_{\rm post}}^{2}{\gamma_{\rm post}}^{2}(1+\gamma)\right.\\
&\left.\quad+\gamma-2x_s v_{\rm post}{\gamma_{\rm post}}^{2}(1+\gamma)\right]=\\
&\rho_{\rm pre}\left[x_s^{2}{\gamma_{\rm pre}}^{2}(1+\gamma)-x_s^{2}\gamma+{v_{\rm pre}}^{2}{\gamma_{\rm pre}}^{2}(1+\gamma)\right.\\
&\left.\quad+\gamma-2x_s v_{\rm pre}{\gamma_{\rm pre}}^{2}(1+\gamma)\right]
\end{aligned}
\end{equation}
where the velocities and Lorentz factors are now measured in the SSS tetrad frame, and $\gamma\equiv c_s^2$ (without subscript) is the squared sound speed of the equation of state $P=\gamma\rho$, not to be confused with the Lorentz factors $\gamma_{\rm pre,post}$. In the limit of a static preshock region ($v_{\rm pre} = 0$), these reduce to the isothermal versions of the jump conditions discussed in \citet{McKee1973}.

For a given postshock state and shock location $x_s$, the two conditions determine the two preshock unknowns $\rho_{\rm pre}$ and $v_{\rm pre}$. Their content is most transparent in the shock rest frame. Writing $w=(v-x_s)/(1-v x_s)$ for the fluid velocity relative to the front and $p\equiv w/\sqrt{1-w^{2}}$ for the corresponding proper velocity, elimination of $\rho_{\rm pre}$ between Eqs.~\ref{eq:jump_sss_n} and~\ref{eq:jump_sss_m} collapses the pair to the single relation
\begin{equation}\label{eq:rh}
p_{\rm pre}\,p_{\rm post} = \frac{\gamma}{1+\gamma},
\end{equation}
the relativistic generalization of the Rankine-Hugoniot condition for isothermal shocks. It follows from the isothermal equation of state and the relativistic baryon- and momentum-flux conditions alone, independent of the self-similar SIS framework. In the non-relativistic limit $p\rightarrow w$ and $\gamma\rightarrow c_s^{2}$, so Eq.~\ref{eq:rh} reduces to the classical relation $w_{\rm pre}w_{\rm post}=c_s^{2}$ \citep{Tsai1995, Shu2002}. Re-expressed in the SSS-frame velocity, Eq.~\ref{eq:rh} becomes a quartic in $v_{\rm pre}$ with up to four real roots: one is the trivial $v_{\rm pre}=v_{\rm post}$ (no discontinuity), and the physical shock is selected from the remainder by the admissibility requirements that the flow be supersonic upstream, subsonic downstream, and compressive across the front. The existence and number of admissible roots, which we exploit in Section~\ref{sec:hydro_shock} to enumerate the exterior solutions, depend on $\gamma$ and on the postshock state.

\subsection{Verification: Newtonian Limit}

As an important consistency check, we verify that Eq.~\ref{eq:jump} recovers known Newtonian results. In the non-relativistic limit, the metric coefficients approach $g_{tt}\rightarrow -1-2\Phi+O(\epsilon^{4})$ and $g_{ij}\rightarrow\delta_{ij}+O(\epsilon^{2})$, where $\epsilon$ is the order of smallness. Defining the standard Newtonian self-similar variables \citep{shu77},
\begin{equation}
\eta=\frac{\zeta}{\sqrt{\gamma}},\quad u(\eta)=\frac{v}{\sqrt{\gamma}},\quad \lambda(\eta)=4\pi t^{2}\rho,
\end{equation}
and taking $\gamma\ll 1$, the GR jump conditions reduce to:
\begin{align}
\lambda_{\rm post}  &= \lambda_{\rm pre}\frac{u_{\rm pre}-\eta_s}{u_{\rm post}-\eta_s},\label{eq:newt_jump1}\\
(u_{\rm post}-\eta_s) &= \frac{1}{u_{\rm pre}-\eta_s}.\label{eq:newt_jump2}
\end{align}
These are identical (up to notational differences) to the jump conditions of \citet{Tsai1995} and \citet{Shu2002}, confirming the algebraic correctness of our relativistic generalization.

\section{Shock Wave Solutions} \label{sec:solutions}
With the jump conditions established and validated against the Newtonian limit, we now apply them to connect the interior collapse solutions to exterior envelopes, constructing the suite of self-similar shock waves that accompany black hole formation. We begin with the simplest case of a hydrostatic envelope before generalizing to dynamic envelopes that infall, expand, or asymptote to rest.

\subsection{Shock Waves with a Hydrostatic Envelope}\label{sec:static_shock}

\begin{figure}
\epsscale{1.0}
\plotone{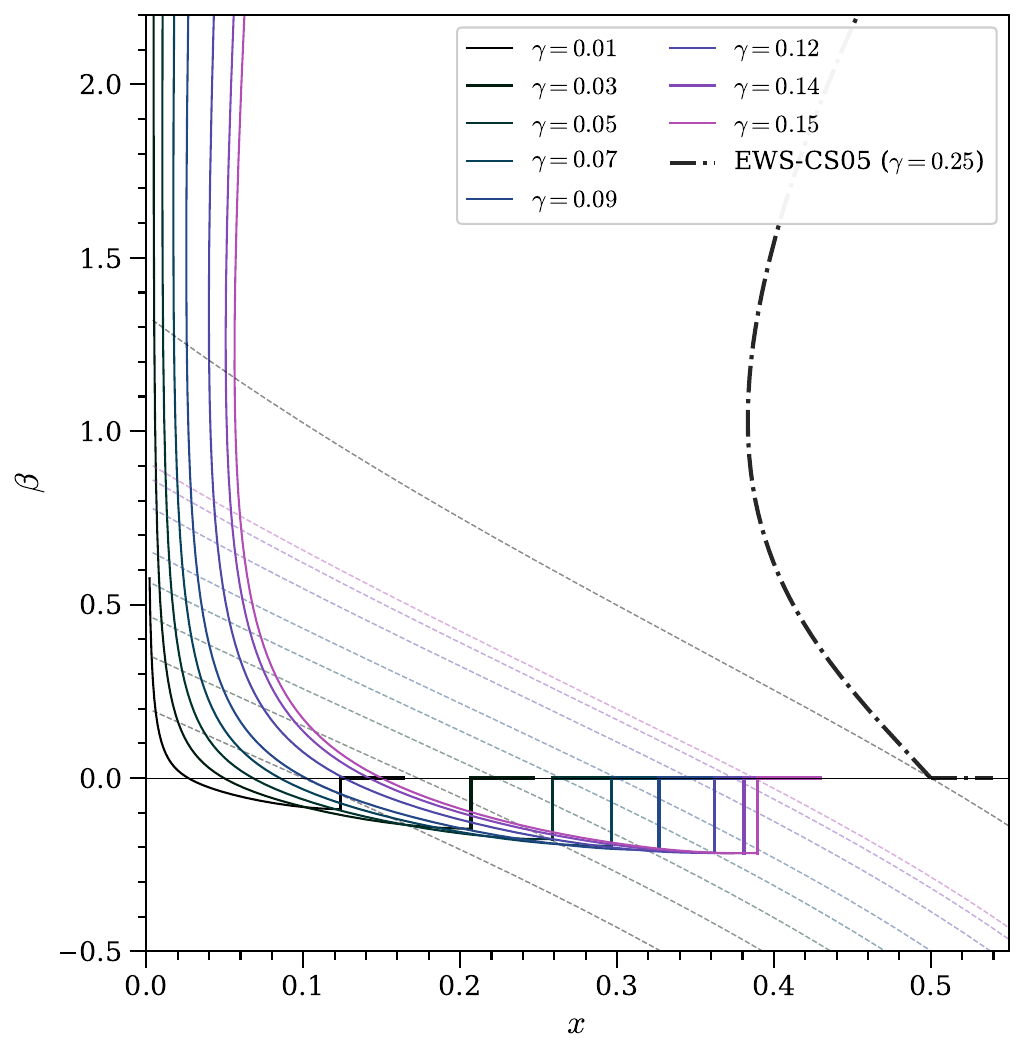}
\caption{Shock wave solutions with a hydrostatic envelope in the $x$--$\beta$ plane. Dashed lines show the sonic critical curves. Solid lines show the complete CSWCP solutions for different $\gamma$ (sound speed squared): each passes smoothly through its critical curve at $\beta_{\rm crit}>0$, with the supersonically collapsing interior extending toward small $x$, crosses $\beta=0$, and connects to the hydrostatic envelope ($\beta=0$) through the shock jump before encountering the critical curve again. The original $\gamma=0.25$ CS05 expansion-wave solution that connects to the static envelope at the critical surface is overlaid (dash-dotted) for comparison.}\label{fig:static}
\end{figure}

We first consider the case where the preshock exterior is in hydrostatic equilibrium ($\beta_0=0$, $\varepsilon_0=1-\Gamma$). The numerical procedure is the following:
\begin{enumerate}
\item Choose a starting point ($\beta_{\rm crit}$, $x_{\rm crit}$) on the critical curve using Eq.~\ref{eq:crit_beta}.
\item Integrate Eq.~\ref{eq:ode} toward increasing $x$.
\item Simultaneously check whether the jump conditions (Eqs.~\ref{eq:jump_sss_n}--\ref{eq:jump_sss_m}) are satisfied between the CSWCP and the hydrostatic exterior.
\item A valid shock exists when both conditions are satisfied and $x_s > \sqrt{\gamma}$ (i.e., the shock is supersonic).
\end{enumerate}

For each value of $\gamma$, we find a unique starting $\beta_{\rm crit}$ that yields a valid shock. Since the two jump conditions determine two unknowns for a fixed preshock boundary condition, the uniqueness of the shock location for each $\beta_{\rm crit}$ is expected.

Results are shown in Figure~\ref{fig:static} and Table~\ref{tab:static}. 
Several notable features emerge. First, the Mach number decreases monotonically as $\gamma$ increases, reaching unity at $\gamma \approx 0.154$, beyond which no shock solution with a hydrostatic envelope exists. This ceiling is a distinctly relativistic phenomenon with no Newtonian analog: since $\gamma \equiv c_s^2$, it corresponds to a sound speed $c_s \approx 0.39\,c$ and a maximum shock velocity of $\sim$40\% of the speed of light, both pressing against the relativistic cap on signal and fluid speeds. In the Newtonian SIS the shock instead propagates at a fixed Mach number for any sound speed \citep{Tsai1995, Lou2004}, so no upper bound on $\gamma$ arises; in the relativistic regime, that same cap forces the upstream Mach number toward unity as $\gamma$ grows, until at $\gamma \approx 0.154$ the conditions for a supersonic shock to exist can no longer be met. Second, in the weakly relativistic limit ($\gamma=0.01$), the Mach number is $\sim$1.24, approaching the Newtonian value of 1.26 found by \citet{Tsai1995}. Third, the existence of a maximum $\gamma$ is physically intuitive: the isothermal assumption requires all shock-generated energy to be radiated away immediately, and the energy cost of accelerating gas to supersonic speeds grows with $\gamma$.

\begin{deluxetable*}{cccccc}
\tablecaption{Shock wave solutions\label{tab:static}}
\tablehead{
\colhead{$\gamma = c_s^2$} & \colhead{$\beta_{\rm crit}$} & \colhead{$v_{\rm shock}$} & \colhead{\tablenotemark{a}$\beta_\infty$} & \colhead{Mach} & \colhead{$\eta = E_{\rm shock}/M_{\rm enc}c^2$}
}

\startdata
\cutinhead{Hydrostatic exterior ($\beta_\infty = 0$; $\gamma \lesssim 0.154$)}
0.01 & 0.190 & 0.124 & 0 & 1.24 & 0.5\% \\
0.03 & 0.331 & 0.207 & 0 & 1.19 & 1.6\% \\
0.05 & 0.430 & 0.259 & 0 & 1.16 & 2.8\% \\
0.07 & 0.512 & 0.297 & 0 & 1.12 & 4.0\% \\
0.09 & 0.582 & 0.327 & 0 & 1.09 & 5.3\% \\
0.12 & 0.675 & 0.362 & 0 & 1.05 & 7.2\% \\
0.14 & 0.730 & 0.385 & 0 & 1.02 & 8.5\% \\
0.15 & 0.756 & 0.394 & 0 & 1.01 & 9.1\% \\
\cutinhead{\tablenotemark{b}Envelope collapse (EC; $\gamma \lesssim 0.155$)}
0.09 & 0.40--0.60 & 0.30--0.34 & $+0.05$ to $+0.66$ & \nodata & 2.6--8.5\% \\
0.12 & \nodata & 0.35--0.37 & $+0.01$ to $+0.69$ & \nodata & 5.1--9.2\% \\
0.14 & \nodata & 0.37--0.39 & $+0.29$ to $+0.59$ & \nodata & 7.7--10.2\% \\
0.15 & \nodata & $\sim$0.39 & $+0.63$ to $+0.67$ & \nodata & 9.1--9.5\% \\
\cutinhead{\tablenotemark{b}Envelope expansion (EE; $\gamma \lesssim 0.13$)}
0.09 & 0.40--0.60 & 0.32--0.35 & $-0.46$ to $-0.02$ & \nodata & 2.8--5.9\% \\
0.12 & \nodata & 0.35--0.36 & $-0.51$ to $-0.14$ & \nodata & 6.6--8.4\% \\
\cutinhead{Breeze\tablenotemark{b} ($\gamma \lesssim 0.13$)}
0.09 & $\sim$0.52 & $\sim$0.35 & $\approx 0$ & \nodata & $\sim$3.4\% \\
\enddata
\tablecomments{$\beta_{\rm crit}$ is the fluid velocity on the critical curve where the CSWCP originates (Eq.~\ref{eq:crit_beta}). $v_{\rm shock}$ is the shock velocity in units of $c$. $\beta_\infty$ is the asymptotic fluid velocity at large $x$ in the preshock exterior ($\beta > 0$ corresponds to infall; Eq.~\ref{eq:e_nd}). $\eta$ is the energy extraction efficiency defined by the energy radiated at the shock front divided by the enclosed rest mass (Eq.~\ref{eq:eta_def}; see Section~\ref{sec:physics}).
For the hydrostatic exterior, all quantities are uniquely determined by the jump conditions; the efficiency rises monotonically to $\approx$9\% at the sonic cutoff $\gamma \approx 0.154$. For hydrodynamic exteriors, a continuous family of solutions exists, parameterized by $\beta_\infty$; the listed $\eta$ gives the range across the admissible solutions, with the overall maximum $\approx$10\% (EC, $\gamma \approx 0.14$).}
\tablenotetext{a}{For EC and EE, their $\beta_\infty$ ranges were obtained from a coarsely sampled numerical survey.}
\tablenotetext{b}{Hydrodynamic exterior solutions. The shock location is not unique---each $\gamma$ admits a continuous family of solutions parameterized by the velocity of the fluid in the pre-shock region $\beta_{\rm pre}$ (which is uniquely characterized by its asymptotic value $\beta_\infty$), subject to the physical admissibility conditions (supersonic upstream, subsonic downstream, compressive). EE and breeze solutions cease to exist for $\gamma \gtrsim 0.13$, where the Rankine-Hugoniot quartic admits no admissible roots with outward or near-zero preshock velocities; EC solutions persist marginally past the hydrostatic limit, to $\gamma \approx 0.155$.}
\end{deluxetable*}

\subsection{Shock Waves with Hydrodynamic Envelopes}\label{sec:hydro_shock}

\begin{figure}[t]
\epsscale{1.0}
\plotone{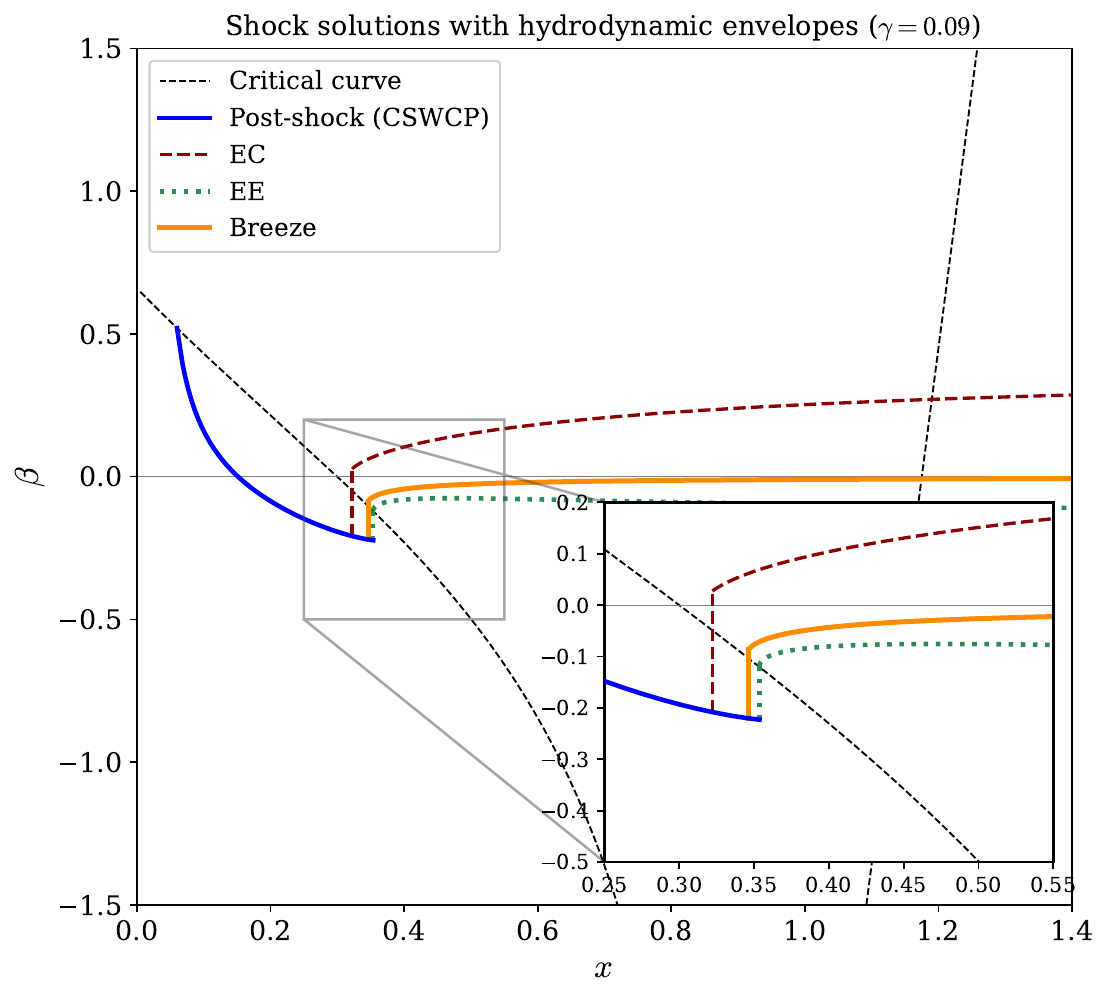}
\caption{Shock solutions with hydrodynamic envelopes for $\gamma=0.09$, showing the critical curve (dashed), the complete CSWCP (blue solid) passing smoothly through the critical curve from the supersonically collapsing interior, and three types of exterior solutions---envelope collapse (EC, red dashed), envelope expansion (EE, green dotted), and a breeze solution (orange solid) approaching $\beta\rightarrow 0$ at large $x$. The inset shows a zoomed view of the shock region near the critical curve.}\label{fig:hydro}
\end{figure}

While the hydrostatic exterior provides a clean demonstration, the spacetime of an SIS is not asymptotically flat. Moreover, in the Newtonian regime, hydrodynamic envelopes play a central role in star formation: the envelope expansion with core collapse (EECC) solutions of \citet{Lou2004} describe protostellar collapse with outflowing envelopes, while the``champagne flow'' solutions of \citet{Shu2002} model the expansion of H~II regions into surrounding molecular clouds. To make the solutions more physically realistic and connect to these well-established Newtonian results, we replace the hydrostatic boundary with three classes of hydrodynamic envelopes:

\begin{enumerate}
\item \textbf{Breeze solution:} The envelope is expanding near the shock front but asymptotically static at large radii ($\beta_0 \rightarrow 0$ as $x\rightarrow\infty$). This is the GR analog of the Newtonian breeze solutions familiar from stellar wind theory and champagne flow models \citep{Shu2002}.

\item \textbf{Envelope expansion (EE):} The exterior has a constant outward velocity at all radii. In the Newtonian limit, this corresponds to the EECC solutions of \citet{Lou2004}.

\item \textbf{Envelope collapse (EC):} The exterior is collapsing everywhere, representing a gravitational catastrophe in which gravity exceeds local pressure at all radii in the preshock region.
\end{enumerate}

The inner CSWCP is obtained identically to the hydrostatic case. At each candidate shock location along the postshock branch, the jump conditions (Eqs.~\ref{eq:jump_sss_n}--\ref{eq:jump_sss_m}) determine the preshock state, subject to the physical admissibility conditions: the upstream flow must be supersonic and the downstream flow subsonic relative to the front, with compression across it. Each admissible preshock state is then integrated outward to large $x$ and classified by its asymptotic envelope velocity, $\beta(x) \rightarrow \beta_\infty$: envelope collapse ($\beta_\infty > 0$), envelope expansion ($\beta_\infty < 0$), or the breeze separatrix ($\beta_\infty = 0$), which matches the asymptotically static power-series envelopes of CS05 (their Eq.~2.18). For the breeze solution, we verify consistency by integrating inward from the large-$x$ power series and recovering the preshock state at the shock.

Representative results for $\gamma = 0.09$ are shown in Figure~\ref{fig:hydro}. At this value of $\gamma$, all three envelope types are available: the admissible shocks form a continuous family in the shock position, along which the asymptotic envelope velocity $\beta_\infty$ decreases monotonically from infall (EC) through the breeze separatrix ($\beta_\infty = 0$) to outflow (EE). However, the availability of exterior types depends strongly on $\gamma$. All three types coexist for $\gamma \lesssim 0.13$; for $\gamma \gtrsim 0.14$, only EC solutions remain---the Rankine-Hugoniot quartic velocity equation admits no admissible roots with outward or near-zero preshock velocities, eliminating the EE and breeze branches entirely. This progressive narrowing of available exterior types with increasing $\gamma$ is likely a consequence of the relativistic kinematics: at higher sound speeds, the jump conditions become increasingly restrictive. For EC exteriors, the shock location is not unique. Unlike the hydrostatic case where the zero-velocity boundary condition uniquely determines $v_{\rm shock}$, the shock solutions exist for a range of $v_{\rm shock}$ that can be connected to EC exteriors parameterized by the preshock infall velocity $\beta_{\rm pre}$ as long as the same physical admissibility conditions (supersonic upstream, subsonic downstream, compressive) are met.

\section{Black Hole Formation and the Comoving-Schwarzschild Bridge} \label{sec:bh}
Since the formation of a black hole would be manifested by the metric coefficients in the SSS formulation becoming singular at the event horizon, we introduce the self-similar comoving (SSC) coordinate of CS05 to describe the formation of the central black hole,
\begin{equation}\label{eq:comoving_metric}
ds^{2}=-e^{2\Phi}dT^{2}+e^{2\Lambda}dR^{2}+e^{2\omega}R^{2}(d{\theta}^{2}+\sin^{2}\theta\, d\phi^{2}),
\end{equation}
where $r=e^{\omega} R$ is the circumferential radius, $\xi=R/T$ is the comoving self-similar variable, and $\Phi$, $\Lambda$, $\omega$ are functions of $\xi$ only. The four-velocity is simply $u^{(a)}=(1,0,0,0)$, and the stress-energy tensor takes the form $T^{(0)(0)}=\rho$, $T^{(i)(j)}=\gamma\rho\delta^{(i)(j)}$, yielding the scaled energy density:
\begin{equation}
\bar{\varepsilon}=4\pi R^{2} (1+\gamma)\rho e^{2\Lambda},
\end{equation}
and the proper self-similar variable $y=e^{\Lambda-\Phi}\xi$.

The simplicity of the comoving stress-energy tensor makes two components of the conservation equation immediately integrable (CS05):
\begin{subequations}\label{eq:comoving_conserv}
\begin{align}
\ln \bar{\varepsilon} &= \frac{2\Gamma\Lambda-4\omega}{1+\Gamma} + C_1 \\
\ln \bar{\varepsilon} &= \frac{2\ln y-2\Gamma\Lambda-2\Gamma \ln \xi}{1-\Gamma} + C_2,
\end{align}
\end{subequations}
where $C_1$ and $C_2$ are integration constants. Combined with the Einstein equations, the system reduces to two independent ODEs in $\bar{\varepsilon}$, $y$, $\Lambda$, and $\Omega\equiv\omega^{\prime}$ (see CS05 for the full expressions).

\subsection{Coordinate Transformation with Zero-Velocity Surface as a Coordinate Bridge}\label{sec:bridge}
In CS05, only the expansion wave solution (EWS) was studied, for which the exterior is in hydrostatic equilibrium with known asymptotic behaviors ($\beta_0 = 0$, $\varepsilon_0 = 1 - \Gamma$). In that case, the static exterior has a closed-form metric, and the comoving and Schwarzschild frames coincide throughout the unperturbed envelope ($\omega = 0$, $\Omega = 0$ where $v = 0$). The known asymptotic values of $\bar{\varepsilon}$, $\Lambda$, and $y$ at large $\xi$ can therefore be substituted directly into Eq.~\ref{eq:comoving_conserv} to fix the integration constants $C_1$ and $C_2$.
For the shock-wave solutions presented here, this simplification is no longer applicable. The shock discontinuity separates the interior collapsing region from the exterior envelope, so the comoving description of the postshock flow cannot be smoothly continued to the preshock boundary---the interior integration constants must be determined independently of the exterior solution. Moreover, for hydrodynamic exteriors (EC, EE, or breeze), the asymptotic behavior is not known in closed form, precluding even the possibility of imposing analytic boundary conditions at large radii. 

For the CSWCP solutions we explored in this work, we found that the SSS and SSC coordinates can still be \emph{uniquely connected} at a specific surface where $\beta=0$ in the SSS, similar to the hydrostatic equilibrium exterior studied in CS05. As shown in Fig. 1, every CSWCP crosses $\beta=0$ at some point in the $x$--$\beta$ plane, separating the inner collapsing region ($\beta>0$) from the outer expanding region ($\beta<0$). At this surface, the three-velocity measured in SSS vanishes: $v=0$. The zero-velocity surface provides the necessary alternative: it is the unique locus where $v = 0$ in SSS and $\Omega = 0$ in SSC simultaneously, so the two coordinate systems coincide, and the SSS solution values can be used to fix $C_1$ and $C_2$ via Eq.~\ref{eq:constants}. We note that the zero-velocity surface necessarily lies outside the event horizon: a vanishing three-velocity presupposes a static observer, which exists only outside the horizon (inside, every trajectory is ingoing toward the singularity). The SSS description is therefore regular at the bridge, so the SSC matching that fixes $C_1$ and $C_2$ is well-defined for every solution we construct.

The comoving four-velocity transforms as:
\begin{equation}
u^{r} = \frac{\partial r}{\partial T}\,u^{T} = -\xi\Omega\, e^{\omega-\Phi}.
\end{equation}
At the zero-velocity surface, $u^{r}=0$ requires $\Omega=0$. Since the SSC frame is defined by vanishing spatial velocity, the condition $u^{r}=0$ in SSS means the Schwarzschild observer is momentarily \emph{comoving} with the fluid. At this event, the two coordinate systems coincide:
\begin{align}
g_{rr} &= g_{RR}, \quad a^{2} = e^{2\Lambda}, \quad x = y, \label{eq:bridge1}\\
\omega &= \beta = \Omega = 0, \quad \varepsilon = \bar{\varepsilon}. \label{eq:bridge2}
\end{align}

This coalescence fixes the integration constants in Eq.~\ref{eq:comoving_conserv}. With $\zeta_c$, $a_c$, $x_c$, and $\varepsilon_c$ denoting the SSS values at the zero-velocity surface:
\begin{equation}\label{eq:constants}
\begin{aligned}
C_1 &= \ln \bar{\varepsilon}_c-(1-\gamma)\Lambda_c, \\
C_2 &= \ln \bar{\varepsilon}_c+\left(\frac{1}{\gamma}-1\right)\Lambda_c+\frac{2}{1-\Gamma}\left( \Gamma\ln \xi_c-\ln y_c\right),
\end{aligned}
\end{equation}
where $\xi_c = \zeta_c$, $\Lambda_c = \ln a_c$, and $y_c = x_c$ by Eqs.~\ref{eq:bridge1}--\ref{eq:bridge2}.

This procedure uniquely determines the SSC solution for each shock wave found in SSS, enabling a complete description of the spacetime from the expanding envelope down to the growing event horizon and central singularity. The general coordinate transformations between the two frames are detailed in Appendix~\ref{app:transform}.

\subsection{Comoving Solutions}\label{sec:comoving_solutions}

\begin{figure*}
\epsscale{1.0}
\plotone{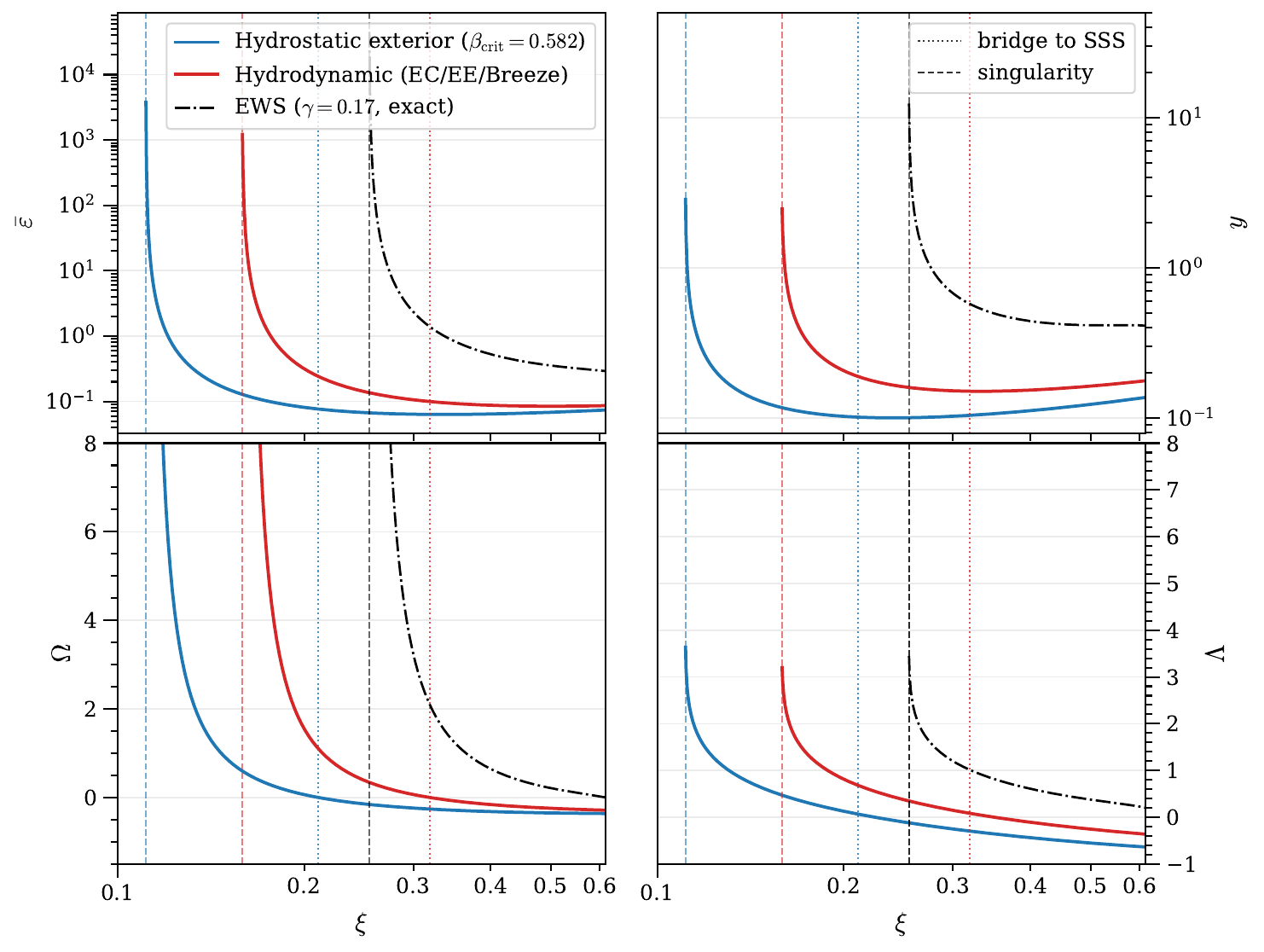}
\caption{Comoving solutions for the shock collapse at $\gamma=0.09$: the scaled energy density $\bar{\varepsilon}$, proper variable $y$, $\Omega=\omega^{\prime}$, and metric coefficient $\Lambda$ as functions of the comoving self-similar variable $\xi$. 
The shock solutions with hydrostatic (blue) and hydrodynamic (red) exteriors are shown. 
Note that different hydrodynamic exteriors share the same comoving interior and differ only in the location of the shock shell $\xi_s \approx 0.83$ (EC), $1.02$ (breeze), and $1.08$ (EE), in units of the common bridge value $\xi_c = \zeta_c$ (equivalently $\xi_s/\xi_c \approx 3.4$, $4.2$, and $4.5$).
Each solution shown here is anchored at its zero-velocity bridge (dotted lines), where the comoving and Schwarzschild frames coincide. 
The divergence of $\bar{\varepsilon}$, $y$, and $\Omega$ at finite $\xi$ (dashed lines) marks the central singularity. 
For comparison, the $\gamma=0.17$ expansion-wave collapse solution (EWS, see CS05) is overlaid (black dash-dotted). 
}\label{fig:comoving}
\end{figure*}

With the integration constants fixed, we numerically integrate the SSC equations outward and inward from the zero-velocity bridge, where the comoving and Schwarzschild descriptions coincide (Eqs.~\ref{eq:bridge1}--\ref{eq:bridge2}): outward to the shock shell, located by $\zeta(\xi)=e^{\omega-\tau}\xi=\zeta_s$, and inward to the center. The combination $E_0=\bar{\varepsilon}\,e^{2\omega}\xi^{\Gamma}/y$ is an exact first integral of the comoving conservation law (Eq.~\ref{eq:comoving_conserv}): it is the weighted product of the two conserved constants, $\ln E_0 = \tfrac{1}{2}(1+\Gamma)\,C_1 + \tfrac{1}{2}(1-\Gamma)\,C_2$, in which the $\Lambda$ dependence cancels, so it is constant along each solution. We verify its conservation to better than $10^{-5}$ along the numerical integration as a consistency check; evaluated at the zero-velocity bridge it gives the constant $E_0$ that later normalizes the enclosed mass (Eq.~\ref{eq:mass}). The comoving state at the shock shell reproduces the postshock SSS state through the transformations of Appendix~\ref{app:transform}. Results for different exterior boundary conditions are shown in Figure~\ref{fig:comoving}.
We find that for each exterior solution explored in this work, a central singularity exists at $\xi=\xi_*$ where $y$, $g_{RR}$, and $|u^{r}|$ diverge,
confirming the formation of a central black hole for the CSWCP solutions and thus the shock wave solutions.

\section{Physical Properties of the Shock Wave} \label{sec:physics}

\subsection{Mass Accretion Rate}
The enclosed mass within radius $R$ is given, following CS05 (their Eq.~5.1), by the volume integral over a comoving $T=\mathrm{const}$ slice of the rest-frame energy density $\rho=T_{\mu\nu}u^{\mu}u^{\nu}$, to which the baryon-number density is proportional for an isothermal gas:
\begin{equation}\label{eq:mass0}
M(R) = \int\rho\sqrt{-g}\,d^{3}x = \frac{1}{1+\gamma}\int^{R}_0 \bar{\varepsilon}\,e^{\Phi-\Lambda+2\omega}\,dR.
\end{equation}
Using the coordinate bridge conditions (Eqs.~\ref{eq:bridge1}--\ref{eq:bridge2}) and the comoving conservation law (Eq.~\ref{eq:comoving_conserv}), this integral can be evaluated in closed form. The enclosed mass within a shell at comoving self-similar variable $\xi_*$ is:
\begin{equation}\label{eq:mass}
M_* = \frac{E_0}{1+\gamma}\,\frac{T_*}{2-\Gamma}\,\xi_*^{2-\Gamma},
\qquad E_0 = \frac{\varepsilon_c\,\zeta_c^{\Gamma}}{x_c},
\end{equation}
where $T_*$ is the elapsed comoving time and the solution constant $E_0$ follows exactly from evaluating the analytic integrals of CS05 (their Eq.~2.25) at the zero-velocity bridge, where $\omega = 0$, $y_c = x_c$, and $\xi_c = \zeta_c$. This expression has the same functional form as the CS05 mass function for the EWS, where the bridge coincides with the critical point and the equilibrium exterior yields $E_0 = (1-\Gamma)/\sqrt{2\gamma}$. For the shock wave solutions, the CSWCP crosses $\beta = 0$ at a different location in the $x$--$\beta$ plane, yielding a bridge energy density $\varepsilon_c < 1 - \Gamma$; we find $\varepsilon_c/(1-\Gamma) \approx 0.4$--$0.5$ across the range of $\gamma$ studied.

We emphasize that $M(R)$ (and the integrated $M_*$) is consistent with the Tolman--Oppenheimer--Volkoff treatment \citep{Tolman1939, OppenheimerVolkoff1939} for the rest mass, which includes the full curved-space volume element, expressed in the self-similar comoving form as $\sqrt{-g}=e^{\Phi+\Lambda+2\omega}R^{2}\sin\theta$. This enclosed mass, which measures the baryonic content of the collapsing core, is distinct from the Misner--Sharp (gravitational) mass $m(r)$, defined by $dm/dr=4\pi r^{2}\rho$ (with $r=e^{\omega}R$ the circumferential radius) and carrying no proper-volume factor \citep{MisnerSharp1964}. The two differ by the gravitational binding energy \citep[e.g.,][]{ShapiroTeukolsky1983}. We normalize the efficiency $\eta$ (Eq.~\ref{eq:eta_def}) to this enclosed mass, following CS05, so that the comparison with accretion radiative efficiencies (e.g.\ the $5.7\%$ Schwarzschild ISCO value) is made on a consistent rest-mass--energy basis.

The normalization $E_0$ of the EWS, $(1-\Gamma)/\sqrt{2\gamma}$, exceeds that of every shock solution, and the EWS singular surface also sits at larger comoving $\xi_\bullet$; numerically, the EWS central accretion rate exceeds that of the hydrostatic-envelope shock solutions by factors of $\sim$5--7. Physically, the shock wave diverts a fraction of the gravitational binding energy into an outward-propagating disturbance, reducing the supply to the central black hole.

Figure~\ref{fig:mdot} shows this comparison across the CSWCP family at $\gamma=0.09$. The central accretion rate is a monotonically decreasing function of $\beta_{\rm crit}$ alone: because the deep interior is causally insulated by its sonic critical point, all envelope types built on the same interior accrete identically---the choice of exterior affects only the location of the shock shell and the energy budget $\eta$ (Eq.~\ref{eq:eta_def})---and the smooth expansion-wave collapse ($\beta_{\rm crit}\rightarrow 0$) bounds the accretion rate from above.

Since $M_*\propto T_*$ in Eq.~\ref{eq:mass}, the dimensionless rate $\dot{M}=E_0\,\xi_*^{2-\Gamma}/[(1+\gamma)(2-\Gamma)]$, evaluated at the singular surface $\xi_*$, is quoted in geometrized units; the corresponding physical rate is $\dot{M}\times c^{3}/G \simeq \dot{M}\times 2.0\times10^{5}\,M_\odot\,{\rm s^{-1}}$. Because the SIS spacetime is not asymptotically flat, the normalization of the self-similar time---and hence the dimensionless value of $\dot{M}$---is conventional; all solutions in Figure~\ref{fig:mdot} share a common normalization (the static-exterior gauge $x=\sqrt{2\gamma}\,\zeta^{\Gamma}$, with each comoving time anchored at the zero-velocity bridge where $T=t$), so that ratios between solutions are convention-free. For a generic interior, which admits a continuum of envelopes, the absolute $\zeta$ scale is fixed by anchoring to the unique admissible envelope with vanishing preshock velocity ($\beta_{\rm pre}=0$), where the static-SIS relation applies exactly; this prescription is exact for the hydrostatic solution ($\beta_{\rm crit}=0.582$, where $\varepsilon_{\rm pre}=1-\Gamma$) and in the EWS limit, and accurate to a few per cent in between. In the Newtonian limit, where the time normalization becomes unambiguous, the construction recovers the classic $\dot{M}=0.975\,a^{3}/G$ of \citet{shu77} for the expansion-wave solution.

\begin{figure*}
\plotone{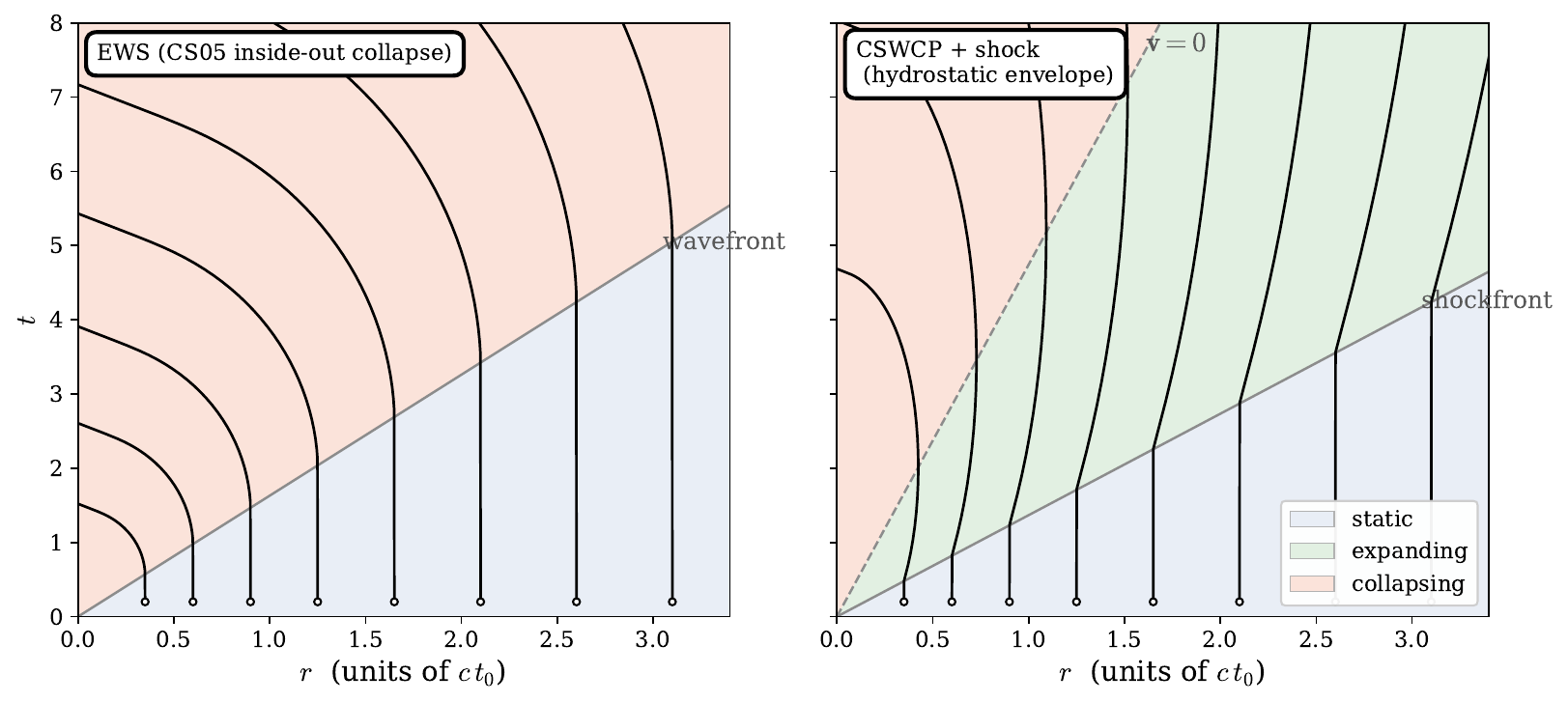}
\caption{
Spacetime $(r, t)$ diagrams contrasting the smooth and shocked collapse solutions. Self-similar structure makes every dynamical feature a surface of constant $\zeta = r/(ct)$, a straight line through the origin. Black curves are fluid worldlines; open circles mark their launch radii at $t_0$. As $t$ increases, each worldline crosses the constant-$\zeta$ feature lines.
Shaded regions denote the local kinematic state of the gas: static ($v=0$, blue), expanding ($v>0$, green), and collapsing ($v<0$, red).
\emph{Left:} the expansion-wave solution (EWS; shown for $\gamma=0.17$, the lowest $\gamma$ with an exact smooth solution). The wavefront (gray line) propagates outward at the sound speed.
\emph{Right:} the CSWCP with a hydrostatic envelope ($\gamma=0.09$). The shock front (gray) is supersonic, and the zero-velocity surface (dashed, $v=0$) lies inside it; fluid worldlines bend across both, tracing the ``fountain'' detour discussed in the text. The two panels are each normalized to their own static SIS envelope and are shown at different $\gamma$, thus the absolute scales between them are not directly comparable.}\label{fig:spacetime}
\end{figure*}

The contrast between the shock wave solutions and the EWS solutions admits a simple Lagrangian interpretation. Because every feature of a self-similar solution sits at fixed $\zeta = r/t$, the shock front, the zero-velocity surface, and the sonic points are \emph{pattern} surfaces that propagate outward through the gas, rather than surfaces locked to the gas.
As a demonstration, we show the space-time diagrams of the EWS and shock solutions with hydrostatic exterior in Fig.~\ref{fig:spacetime}.
In the EWS, a fluid shell remains static until the expansion-wave front arrives, and then falls monotonically onto the black hole: every shell makes a single inbound trip. In a shock wave solution, a shell of the envelope is instead struck by the supersonic front and driven \emph{outward}; it decelerates in the gravitational field, stalls just as the outward-propagating zero-velocity surface overtakes it, and only then reverses and falls in, re-accelerating through the interior sonic point onto the black hole. Each shell therefore makes a detour---a fountain---before collapsing. Between the shock front and the zero-velocity surface, the flow is locally a decelerating self-gravitating outflow; its smooth re-crossing of the critical curve is forbidden by the second law of thermodynamics, which is why the discontinuity is required in the first place (Section~\ref{sec:jump}). This detour is the common origin of the two effects quantified in this section: the reversal at the front dissipates the relative kinetic energy of the gas crossing it, which the isothermal condition converts into the radiated energy quantified by $\eta$ (Eq.~\ref{eq:eta_def}), while the round trip throttles the rate at which shells reach the center, suppressing $\dot{M}$ by the factor of $\sim$5--7 compared to the EWS limit shown in Figure~\ref{fig:mdot}. The deep interior, causally insulated inside its sonic critical point, collapses in the same manner in either case. 

\begin{figure}
\epsscale{1.2}
\plotone{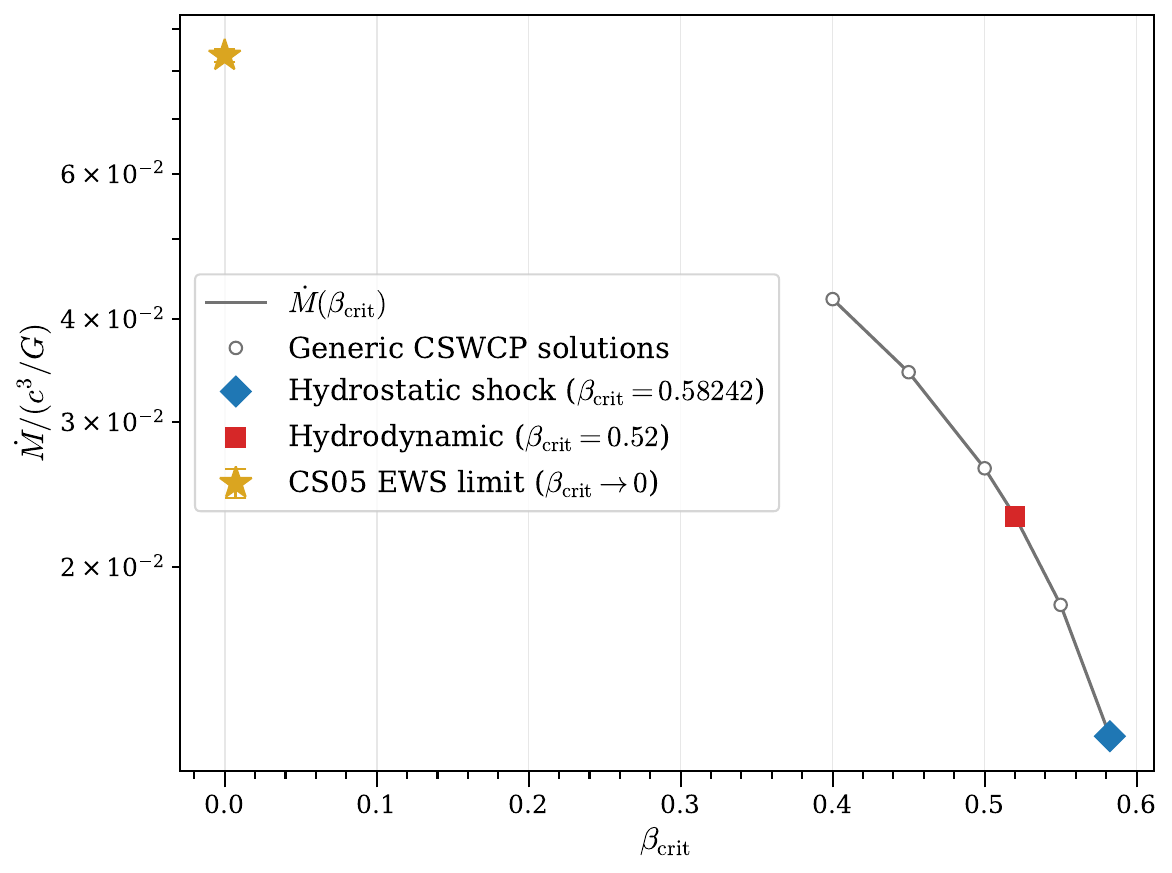}
\caption{Central accretion rate $\dot{M}$ (units of $c^{3}/G$) versus the interior parameter $\beta_{\rm crit}$ at $\gamma=0.09$, in the common gauge described in the text. Symbols: the hydrostatic-exterior shock solution (diamond), the hydrodynamic-family interior of Figures~\ref{fig:hydro} and~\ref{fig:comoving} (square; a single point for the EC, breeze, and EE members, which share this interior), and the EWS limit $\beta_{\rm crit}\rightarrow 0$ (star). CSWCP solutions integrated from a range of $\beta_{\rm crit}$ are also shown as circles. All solutions lie on a single curve $\dot{M}(\beta_{\rm crit})$: the accretion rate is set by the interior alone, and the shock reduces the mass supply to the black hole by a factor of $\sim$7 relative to the smooth expansion-wave collapse.}\label{fig:mdot}
\end{figure}

\subsection{Energy Release at the Shock Front}

The energy radiated at the shock front is evaluated from the component of stress-energy conservation that was \emph{not} imposed as a jump condition---the energy flux. The energy crossing the moving front per unit coordinate time is the jump of the fluid $t$-energy flux through the shock surface, $\dot{E}_{\rm shock} = \Delta\big[4\pi r^{2}\alpha a\,(T^{r}{}_{t}-\zeta_s T^{t}{}_{t})\big]_{\rm pre-post}$, where the full Jacobian of the SSS metric, $\sqrt{-g}=\alpha a r^{2}\sin\theta$, must be retained. Using $\alpha/a = \zeta/x$, this evaluates to
\begin{equation}\label{eq:energy_flux}
\dot{E}_{\rm shock} = \Delta\left[\varepsilon\,\frac{\zeta_s^{2}}{x_s^{2}}\left(\gamma_{L}^{2}\,(x_s - v) - \frac{x_s\,\gamma}{1+\gamma}\right)\right]_{\rm pre-post},
\end{equation}
where $v$ and $\gamma_{L}=(1-v^{2})^{-1/2}$ are the fluid velocity and Lorentz factor measured by the local static observer. We caution that the flat-space shortcut $\sqrt{-g}=r^{2}\sin\theta$, while admissible for the \emph{local} jump conditions by the equivalence principle, is not adequate for the global energy bookkeeping: the lapse and radial metric factors weight the flux when it is accumulated over coordinate time, and omitting them renders the efficiency dependent on the arbitrary self-similar normalization.

To define an energy extraction efficiency, we compare the total energy released over the collapse history to the enclosed baryonic mass at the shock front, using Eq.~\ref{eq:mass} with $\xi_*=\xi_s$:
\begin{equation}\label{eq:eta_def}
\eta \equiv \frac{\dot{E}_{\rm shock}\,T_s}{M_{\rm enc}} = \dot{E}_{\rm shock}\,e^{\tau_s}\,\frac{(1+\gamma)(2-\Gamma)}{E_0\,\xi_s^{2-\Gamma}},
\end{equation}
where $e^{\tau_s}$ maps the SSS coordinate time to the comoving time at the front ($t = e^{\tau}T$; Appendix~\ref{app:transform}), obtained from the bridge-anchored comoving integration. Defined this way, $\eta$ is invariant under the self-similar time-gauge freedom $\zeta \rightarrow \lambda\zeta$, and it is robust to the choice of slicing: an independent computation using SSS constant-$t$ slices with the matching lapse-weighted enclosed mass agrees with Eq.~\ref{eq:eta_def} to within $\sim$5\% for $\gamma \lesssim 0.1$ and $\sim$7\% at the highest $\gamma$ studied, the residual reflecting the dependence of the energy-to-enclosed-mass ratio on the spacelike foliation, which grows with the relativistic strength of the flow.


\begin{figure}
\epsscale{1.2}
\plotone{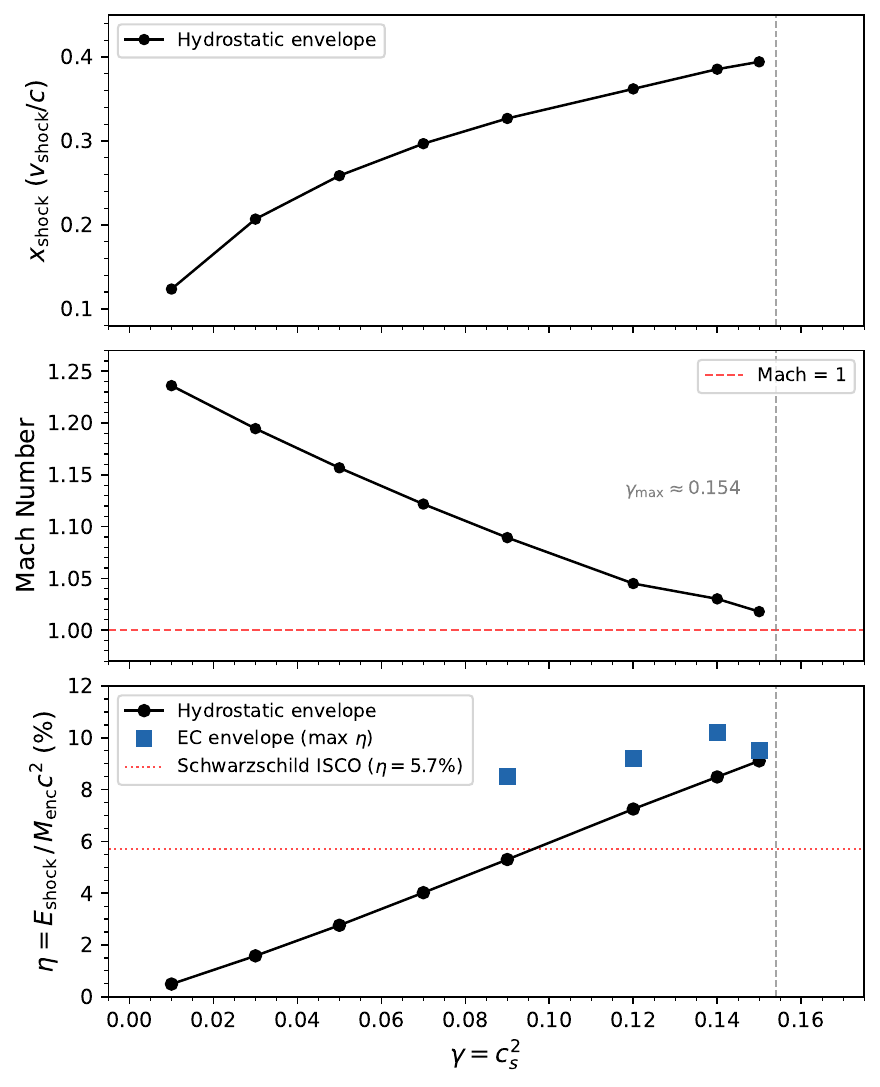}
\caption{Shock wave properties as a function of sound speed $\gamma = c_s^2$. Top: shock velocity $v_{\rm shock}$ in units of $c$. Middle: Mach number $v_{\rm shock}/\sqrt{\gamma}$, showing the approach to unity at the static existence limit $\gamma \approx 0.154$. Bottom: energy extraction efficiency $\eta = E_{\rm shock}/M_{\rm enc}c^2$ for the hydrostatic solutions (curve), the maxima of the EC families (points), and the Schwarzschild ISCO efficiency of 5.7\% (dotted line; see Table~\ref{tab:static}).}\label{fig:shock_properties}
\end{figure}

Results are listed in Table~\ref{tab:static} and illustrated in Figure~\ref{fig:shock_properties}. Supersonic shock solutions with a \emph{hydrostatic} envelope cease to exist beyond $\gamma\approx 0.154$, where the shock becomes sonic (Table~\ref{tab:static}). Because the preshock gas of an envelope-collapse (EC) exterior is already infalling, the upstream Mach condition is relaxed and EC solutions remain available marginally beyond the hydrostatic boundary, to $\gamma \approx 0.155$; envelope-expansion and breeze exteriors, whose preshock gas moves outward, cease to exist already at $\gamma \approx 0.13$. The hydrostatic efficiency rises monotonically with $\gamma$, from $\sim$0.5\% at $\gamma = 0.01$ to $\sim$9\% at the sonic cutoff; across the hydrodynamic families the maximum is $\eta \approx 10\%$ (EC, $\gamma \approx 0.14$). The overall range of efficiencies across all $\gamma$ and exterior types is therefore $\sim$0.5\% to $\sim$10\%---up to $\sim$1.8 times the maximum radiative efficiency for accretion onto a non-spinning (Schwarzschild) black hole, 5.7\% at the innermost stable circular orbit \citep{Misner1973}. The shock mechanism thus taps directly into the gravitational binding energy liberated by the changing spacetime geometry during collapse, providing an energy extraction channel that can exceed accretion in efficiency.

The efficiency varies across each category of hydrodynamic solutions: the maximum $\eta$ occurs for the stronger-infall EC members (smaller $x_s$ and larger $\beta_\infty$), not for those closest to the hydrostatic limit. At $\gamma = 0.09$ the admissible EC efficiencies span $\sim$2.6--8.5\%, rising to the categorical maximum of $\sim$10.2\% at $\gamma = 0.14$.

\section{Discussion and Astrophysical Implications} \label{sec:discussion}

We have derived a suite of self-similar shock-wave solutions for the general-relativistic collapse of a singular isothermal sphere accompanied by the formation of a central black hole. The key mathematical results are: 
\begin{enumerate}
    \item the derivation of GR jump conditions for an isothermal fluid, verified against the Newtonian limit;
    \item the identification of shock wave solutions with static, expanding, and collapsing envelopes;
    \item a coordinate-matching technique using the zero-velocity surface of the collapse solutions to bridge the Schwarzschild and comoving descriptions, which shows that the central accretion rate is set by the interior collapse alone and is suppressed by a factor of $\sim$5--7 relative to the shock-free expansion-wave solution from CS05; and
    \item an energy extraction efficiency of up to $\sim$10\% of the enclosed rest mass, nearly twice the radiative efficiency of Schwarzschild accretion. 
\end{enumerate}   
These results provide an analytical framework for understanding the energetics of black hole formation with significant implications for SMBH seed assembly and relativistic transients.

\subsection{Origin and Nature of the GR Shock Wave}

Unlike Newtonian shock waves in collapsing isothermal spheres, which arise from the interaction of the collapse front with the initial density perturbation, the GR shock wave is fundamentally driven by the dynamical change in spacetime geometry accompanying black hole formation. The collapsing interior produces a growing event horizon whose influence propagates outward. Within the self-similar framework, this manifests as the inability of collapse solutions to smoothly rejoin the exterior without violating the second law of thermodynamics, necessitating a shock discontinuity.

The zero-velocity surface---where $\beta=0$ in SSS and $\Omega=0$ in SSC---plays a dual role. Physically, it separates the infalling gas from the shock-accelerated outflow. Mathematically, it provides the unique bridge between the Schwarzschild and comoving self-similar coordinates, enabling a complete description of the spacetime from the expanding envelope to the growing singularity.

\subsection{Implications for SMBH Seed Formation}

The energy extraction efficiency of up to $\sim$10\% of the enclosed rest mass has significant implications for models of early black hole growth. In the direct collapse scenario \citep{Loeb_Rasio_1994, BL01, Lodato_Natarajan_2006, Begelman2006, LatifFerrara2016}, an atomic-cooling halo with a baryonic mass of $M\sim 10^{4}$--$10^{5} \Msun$ collapses to form a black hole seed, potentially facilitated by shock-heated gas from protogalactic collisions \citep{Inayoshi2015}. The gas needs to be metal-free cooling; Lyman-Werner radiation at a sufficiently high level should also be present in the background to destroy hydrogen molecules and prevent fragmentation (see, e.g., the review by \citealt{Inayoshi_2025_review}).

If the collapse proceeds through an isothermal phase and generates a shock wave of the type described here, the energy release would be:
\begin{equation}
E_{\rm shock} \lesssim 0.1\,M_{\rm enc}\,c^2 \sim 10^{57}\text{--}10^{58}\;\text{erg}
\end{equation}
for an enclosed mass of $10^{4}$--$10^{5}\,M_\odot$. 

An interesting implication is that the shock energy release occurs at the moment of black hole formation itself, before any prolonged phase of radiatively efficient accretion. Most feedback models for early SMBH growth assume that the dominant energy source is subsequent accretion onto the seed black hole. In contrast, the mechanism described here injects energy directly during the collapse process. Even if only a fraction of the shock energy couples to the surrounding gas, the resulting pressure support could temporarily inflate the collapsing envelope, reduce the instantaneous inflow rate, and establish a self-regulated growth phase. Such a mechanism may help explain how massive seeds avoid catastrophic overfeeding while remaining embedded within dense gas reservoirs capable of sustaining later accretion.

This energy injection into the surrounding medium could have several consequences. It could provide radiative feedback that regulates subsequent accretion, potentially contributing to the observed black hole--host galaxy mass correlations even at the earliest epochs \citep{Pacucci_2023_JWST, Gupta_2026}. It could also produce an electromagnetic transient detectable by future surveys, serving as a direct observational signature of heavy seed formation \citep{Pacucci_2026_DCBH}.

At the same time, the shock solutions predict accretion rates suppressed by factors of a few relative to the smooth expansion-wave collapse solution. This suggests that black hole formation may involve an initial feedback-regulated phase in which part of the available gravitational binding energy is redirected into the surrounding medium rather than into the central object. The competition between shock-powered energy deposition and continued infall may therefore influence both the final seed mass and the timescale over which the system emerges as an observable accreting SMBH \citep{Pacucci_2026_DCBH}.

Recent JWST observations have revealed an unexpectedly abundant population of active black holes at $z > 4$, including ``little red dots''---compact, red sources \citep{Matthee2024} now interpreted as young SMBHs of $\sim 10^{5}$--$10^{7}\,M_\odot$ enshrouded in dense ionized cocoons \citep{Natarajan2024, Pacucci_Hernquist_2025, Pacucci_2026_DCBH, Naidu_2025a, Rusakov2026}. The direct collapse pathway and the late-stage collapse of quasi-stars \citep{Begelman2026} have been invoked to explain these heavy seeds, but the energetics and dynamics of the collapse itself are poorly constrained. 

This picture is particularly intriguing in light of the emerging interpretation of little red dots as rapidly growing black holes embedded in optically thick gaseous cocoons. The shock solutions found here naturally produce a configuration in which a substantial fraction of the liberated energy is deposited into the gas surrounding the nascent black hole rather than being advected directly into the horizon. The resulting heated envelope could increase the cocoon's scale height, enhance obscuration, and delay the dispersal of the surrounding gas. In this sense, the shock provides a physically motivated mechanism for creating and maintaining the dense environments inferred around young SMBHs at high redshifts.

Our self-similar shock solutions provide a first-principles estimate of the energy budget available during the formation of such seeds. If the collapse proceeds through an isothermal phase, the $\sim$0.5--10\% energy extraction efficiency derived here implies that a notable fraction of the rest mass energy is deposited into the surrounding cocoon, potentially powering the observed emission from little red dots and contributing to the expansion of their dense envelopes.

In general, the shock wave may represent a previously overlooked stage of direct collapse black hole formation: an intermediate phase between dynamical collapse and sustained accretion, during which the newly formed black hole remains deeply buried inside a shock-heated, radiation-supported envelope. Such a phase would be difficult to observe directly, but could provide a natural evolutionary link between direct collapse seed formation and the obscured SMBH populations now being uncovered by JWST.

\subsection{Connection to Gamma-Ray Burst Energetics}

The collapsar model for long-duration gamma-ray bursts involves the collapse of a massive stellar core to a black hole, with a relativistic jet powered by accretion \citep{Woosley1993, MacFadyen1999}. While the GRB jet is typically attributed to magnetically driven outflows or neutrino annihilation near the black hole, our results suggest an additional energy source: the shock wave generated by the spacetime transition during black hole formation itself.

For a collapsing core with $M\sim 10\,M_\odot$ and an effective sound speed $\gamma \sim 0.1$, the shock energy would be $E_{\rm shock}\sim 0.05\,Mc^2 \sim 10^{54}$ erg---within an order of magnitude of the total energy budget inferred for GRBs. While the isothermal assumption is an idealization for stellar cores, the qualitative result---that several percent of the rest mass energy is channeled into an outward-propagating shock---suggests that gravitational energy release during black hole formation may contribute to the energetics of relativistic transients, complementing the Blandford-Znajek mechanism \citep{BlandfordZnajek1977} and neutrino-driven processes.

\subsection{Caveats and Future Directions}

The principal limitations of this work stem from the two defining assumptions of the SIS model: an isothermal equation of state and spherical symmetry. The isothermal condition ($P = \gamma\rho$) requires that all thermal energy generated at the shock front be radiated away instantaneously, so that the pre- and postshock gas share the same effective temperature. While this is a reasonable first approximation for optically thin, efficiently cooled environments---and is the standard assumption in the Newtonian SIS literature \citep{shu77, Tsai1995, Lou2004}---real astrophysical collapse involves radiation transfer, nucleosynthesis, and non-equilibrium thermodynamics that can produce temperature gradients. 

The assumption of spherical symmetry precludes angular momentum, magnetic fields, and non-radial instabilities, all of which are expected to play a role in realistic collapse scenarios. Despite these idealizations, the SIS framework retains considerable value: it is one of the few settings in which the full nonlinear dynamics of general-relativistic gravitational collapse can be solved analytically, yielding closed-form expressions for the shock velocity, energy release, and mass accretion rate. The solutions presented here therefore offer an alternative analytical description of black hole formation that complements numerical relativity simulations by offering analytical benchmarks for testing GRHD shock-capturing schemes \citep{Font2003}, while serving as a baseline against which the effects of more realistic physics can be assessed.

Future work should address several directions. First, relaxing the isothermal assumption to polytropic or more general equations of state would broaden applicability, extending previous pseudo-Newtonian treatments \citep{Lian2014} into the full general-relativistic domain. Second, the inclusion of rotation and magnetic fields---building on the magnetized EWS of CS05---could connect more directly to collapsar models. Third, numerical relativity simulations could test whether the self-similar shock wave solutions identified here emerge as attractors in the full nonlinear evolution. We note that our solutions are spherically symmetric and therefore do not emit gravitational waves by Birkhoff's theorem. However, realistic collapse scenarios will inevitably break spherical symmetry through rotation, turbulence, or fragmentation. In such cases, the rapid rearrangement of the mass distribution associated with the shock could produce a gravitational wave signature potentially detectable by future space-based interferometers such as LISA for SMBH seed formation events.



\begin{acknowledgments}
C.-T.J.C. gratefully acknowledges the late Frank H.S. Shu for his helpful suggestions and discussions, and thanks Fred Adams for insightful feedback that improved the manuscript. C.-T.J.C. acknowledges NASA funding from cooperative agreement 80NSSC24M0035. F.P. acknowledges support from NASA through Chandra Award Nos. DD4-25147X and GO3-24087A, and through JWST Award Nos. JWST-GO-03805.010-A and JWST-GO-05407.015-A. These awards are administered by the Chandra X-ray Center and the Space Telescope Science Institute, respectively. 
\end{acknowledgments}

\appendix

\section{Coordinate Transformation Details}\label{app:transform}

The general coordinate transformation between SSS and SSC follows from $r=e^{\omega} R$ and $t=e^{\tau} T$, with $\zeta=\xi e^{\omega-\tau}$. The metric components transform as:
\begin{align}
g^{rr} &= e^{2\omega-2\Lambda} (1+\Omega)^{2}-e^{2\omega-2\Phi} \xi^{2}\Omega^{2}\\
g^{tt} &= e^{2\tau-2\Lambda}\xi^{-2}\tau'^{2}-e^{2\tau-2\Phi}(1-\tau')^{2}\\
g^{rt} &= e^{\omega+\tau-2\Lambda}\xi^{-1}\tau'(1+\Omega)+e^{\omega+\tau-2\Phi} \xi\Omega(1-\tau')=0,
\end{align}
where the orthogonality of $r$ and $t$ ($g^{rt}=0$) determines:
\begin{equation}
\tau'=-\frac{y^{2}\Omega}{(1+\Omega)-y^{2}\Omega}.
\end{equation}

The four-velocity components transform as:
\begin{align}
u^{r} &=-\xi\Omega\, e^{\omega-\Phi},\\
u^{t} &=e^{\tau-\Phi}(1-\tau').
\end{align}

Using $\bar{\varepsilon}=e^{2\Lambda-2\omega}\varepsilon/a^{2}$ and the substitutions $\xi=e^{\Phi-\Lambda}y$ and $e^{2\omega-2\Lambda}=\varepsilon/(\bar{\varepsilon}a^{2})$, the transformations in terms of scaled variables become:
\begin{align}
u^{r} &: \quad -\sqrt{\frac{\varepsilon}{\bar{\varepsilon}}}\,y\Omega=\frac{v}{\sqrt{1-v^{2}}} \label{eq:trans_ur}\\
u^{t} &: \quad \frac{\varepsilon}{\bar{\varepsilon}}(1-\tau')^{2}y^{2}=\frac{x^{2}}{1-v^{2}} \\
g^{rr} &: \quad \frac{1}{a^{2}}=\frac{\varepsilon}{\bar{\varepsilon}a^{2}}\left[(1+\Omega)^{2}-y^{2}\Omega^{2}\right] \\
g^{tt} &: \quad x^{2}=y^{2}\frac{\varepsilon}{\bar{\varepsilon}}\left(\frac{\tau'}{y^{2}}-(1-\tau')^{2}\right).
\end{align}

\section{Newtonian Limit of the Governing Equations}\label{app:newtonian}

For completeness, we show that the GR equations recover the standard Newtonian SIS equations in the appropriate limit. With the metric coefficients $g_{tt}=-1-2\Phi+O(\epsilon^{4})$ and $g_{ij}=\delta_{ij}+O(\epsilon^{2})$, and defining:
\begin{align}
\beta &= -2\sqrt{\gamma}\,u, \quad \varepsilon = \lambda\gamma\eta^{2}, \\
a^{2} &= 1+O(\epsilon^2), \quad x = \sqrt{\gamma}\,\eta\left(1-\Phi+O(\epsilon^2)\right),
\end{align}
the self-similar Poisson equation gives:
\begin{equation}
\frac{d\Phi}{d\eta}=\lambda\gamma(\eta-u),
\end{equation}
and defining the Newtonian mass function $M(r,t)=\gamma^{3/2}t\,m(\eta)$, the continuity equation reduces to $m=\eta^{2}\lambda(\eta-u)$. The fluid equations become:
\begin{align}
\frac{du}{d\eta} &= \left[\lambda(\eta-u)-\frac{2}{\eta}\right]\frac{\eta-u}{(\eta-u)^2-1}, \\
\frac{d\log\lambda}{d\eta} &= \left[\lambda-\frac{2}{\eta}(\eta-u)\right]\frac{\eta-u}{(\eta-u)^2-1},
\end{align}
which are identical to the Newtonian SIS equations of \citet{shu77}.

\bibliography{references}
\bibliographystyle{aasjournal}

\end{document}